\documentclass{pasj00}
\draft

\begin{document}
\SetRunningHead{Author(s) in page-head}{Running Head}
\Received{2016/10/11}
\Accepted{2017/01/19}
%\Published{}%{yyyy/mm/dd}

\title{Black hole spin of Cyg~X-1 determined from the softest state ever observed}

%%% begin:list of authors
% Do NOT capitalize all letters in "textsc".
\author{Takafumi \textsc{Kawano}$^{1}$, Chris \textsc{Done}$^{2,3}$, Shin'ya {\sc Yamada}$^{4}$,  
Hiromitsu {\sc Takahashi}$^{1}$, Magnus {\sc Axelsson}$^{4}$, and Yasushi {\sc Fukazawa}$^{1}$
%\thanks{Last update: January 19, 2007}
}
\affil{$^{1}$Department of Physical Science, Hiroshima University,
1-3-1 Kagamiyama, Higashi-Hiroshima, Hiroshima 739-8526, Japan}
\affil{$^{2}$Centre for Extragalactic Astronomy, Department of Physics, University of Durham, Durham, UK}
\affil{$^{3}$Institute of Space and Astronautical Science (ISAS), Japan Aerospace Exploration Agency (JAXA), Kanagawa, Japan }
\affil{$^{4}$Department of Physics, Tokyo Metropolitan University, Tokyo, Japan}

\email{tkawano@hep01.hepl.hiroshima-u.ac.jp}
%%% end:list of authors

%%% Please use the following style in case that sorting by 
%%% affiliation is impossible. 
%
% \author{%
%   D-Firstname \textsc{D-Familyname}\altaffilmark{1}
%   E-Firstname \textsc{E-Familyname}\altaffilmark{1,2}
%   and
%   F-Firstname \textsc{F-Familyname}\altaffilmark{2}}
% \altaffiltext{1}{Address of Institute}
% \email{ddddd@xxx.xxx.xx.xx}
% \email{eeeee@xxx.xxx.xx.xx}
% \altaffiltext{2}{Address of Institute}

%% `\KeyWords{}' always has to be placed before `\maketitle'.
%\KeyWords{xxxx:xxxx ......} %Do NOT move this preamble from here!
%\KeyWords{black hole${}_1$ --- X-ray binaries${}_2$ --- Cyg~X-1${}_3$}

\maketitle

\begin{abstract}

We show the softest ever spectrum from Cyg~X-1, 
detected in 2013 with Suzaku.  
This has the weakest high energy Compton tail ever seen from this object, 
so should give the cleanest view of the underlying disk spectrum,
and hence the best determination of black hole spin from disk continuum fitting. 
Using the standard model of a disk with simple non-thermal Comptonisation 
to produce the weak high energy tail gives a high spin black hole. 
However, we get a significantly better fit by including an additional, 
low temperature thermal Comptonisation component, 
which allows a much lower black hole spin. 
Corroboration of the existence of an additional Compton component 
comes from the frequency dependent hard lags 
seen in the rapid variability in archival high/soft state data. 
These can not be explained if the continuum is a single non-thermal Comptonisation component, 
but are instead consistent with a radially stratified, 
multi zone Comptonisation spectrum, 
where the spectrum is softer further from the black hole. 
A complex multi-zone Comptonisation continuum is required 
to explain both spectra and timing together, 
and this has an impact on the derived black hole spin. 

\end{abstract}

\section{Introduction}\label{sec:intro}

Cyg~X-1 was the first black hole candidate to be identified, 
so is one of the best studied. 
However, 
it has some differences compared to the majority of black hole binaries in our Galaxy.  
It is a persistent rather than transient system, 
showing that the outer disk is always above the hydrogen ionization instability 
(\cite{van1996}). 
However, even though the system has a high mass companion star,
the mass accretion rate from this is probably not high enough 
for the disk to remain stable in purely Roche lobe overflow, 
so the disk may be partly wind fed (\cite{Coriat2012}).

While the resulting mass accretion rate onto the black hole is rather stable, 
its value is close to that of dramatic spectral transition, 
so the system shows strong spectral variability. 
It switches between the Compton dominated low/hard state 
towards the disk dominated high/soft state. 
This is probably the result of a change in the nature 
of the accretion flow from a hot, optically thin, 
geometrical thick (advection dominated) accretion flow to a cool, optically thick, 
geometrically thin disk (see e.g. the review \cite{Done2007}). 
In Cyg X-1 this spectral switch occurs across 
only a factor 2-3 change in overall luminosity (\cite{Zdziarski2002}).
This is very different to the transient systems, 
where the mass accretion rate rises dramatically from quiescence to outburst, 
and the state transition takes place against a background 
of a strongly increasing mass accretion rate (e.g. the compilation of \cite{Dunn2010}).  
This may be the origin of the differences seen 
in the luminosity of the hard-to-soft spectral state change 
between Cyg~X-1 and the transients. 
The transients show large scale hysteresis, 
where the hard-to-soft transition 
during the rapid outburst rise can occur at a much higher 
(up to at least a factor of 10) luminosity 
than the reverse transition on the decline, 
and where the luminosity of the hard-to-soft transition itself is variable 
between different outbursts of the same source 
(e.g. \cite{Smith2002, Maccarone2003, Gierlinski2006, Gladstone2007, Yu2009, Dunn2010}).
By contrast, in Cyg~X-1, the hysteresis is only small scale, 
less than a factor of 2-3, 
and the hard-to-soft transition luminosity is fairly stable 
(\cite{Smith2002, Zdziarski2002, Meyer2007}). 

The transients often show high/soft spectra 
which are almost completely dominated by the disk, 
and where the disk peak temperature and luminosity vary together 
as expected for a constant inner disk radius as the mass accretion rate changes. 
This gives confidence that this radius is a fundamental property of the system, 
as expected if this is the innermost stable circular orbit. 
This opens a way to measure black hole spin 
if the system parameters are known accurately 
(e.g. the reviews by \cite{Remillard2006, Done2007}).  
However, in Cyg~X-1 the high/soft state spectra 
always have a fairly strong tail of emission to high energies (e.g. \cite{Negoro2010}), 
showing that there is a non-negligible fraction of the accretion power 
which is not dissipated in the thin disk. 
This could mean that Cyg~X-1 never quite makes a full transition, 
so that the accretion disk never quite reaches down to the last stable circular orbit, 
and/or that the coronal emission is fed directly 
from the stellar wind rather than via the accretion disk (\cite{Sugimoto2016}). 
This latter idea might explain its rather different X-ray variability power spectra 
compared to high/soft states with similarly strong tails in the transients (\cite{Done2005}). 

Whatever the origin, the strong tail in the high/soft state of Cyg~X-1,
it makes the spectra difficult to model as it has complex curvature. 
Early fits to 0.7-200~keV broadband 
data from Cyg X-1 required an additional low temperature thermal Comptonisation component 
as well as the more usual high/soft state components of a disk, 
high energy Compton tail and its reflection from the disk (\cite{Gierlinski1999, Zdziarski2002}). 
However, while these fits used sophisticated models 
for the Comptonisation (with both thermal and non-thermal electrons),
they used rather simple models of the accretion disk
spectra which assumed zero spin.
Instead, more recent fits have used some of the best current models 
for the disk emission in full general relativity, 
but only included simple Comptonisation models, 
where the electrons have a single non-thermal power law distribution 
(\cite{Gou2011, Gou2014, Tomsick2014, Walton2016}). 
These fits require a very high spin black hole, 
in sharp contrast to the earlier fits with thermal and non-thermal Comptonisation 
which fit the data with a zero spin black hole disk model. 
Unlike the transients, 
Cyg X-1 does not show a wide range in luminosity for its high/soft state, 
so the apparent disk radius cannot be compared across different mass accretion rates to see
if it remains constant as predicted by the high spin solution.

Instead, another way to tell the difference between 
these two very different physical models is fast time variability. 
The disk does not vary on timescales of less than a few seconds 
in the high/soft state (e.g. \cite{Churazov2001}), 
so variability on faster timescales must instead be connected 
to the Comptonization region(s). 
This fast variability shows that the light curves at 
higher energies are correlated with those at lower energies, 
but with a time lag, and this lag gets shorter for faster variability timescales
(\cite{Miyamoto1989, Nowak1999, Pottschmidt2000, Grinberg2014}). 
Such frequency dependent time lags are commonly
seen in the low/hard state, 
where they are now interpreted as fluctuations propagating down 
through a Comptonising flow which is radially stratified in both spectrum and variability. 
Slow fluctuations are stirred up at larger radii, where the spectra are softer.
These propagate down to modulate the emission from smaller radii 
where the spectra are harder, 
giving rise to a lag of the hard band relative to the soft.
Faster fluctuations are stirred up closer to the black hole, 
so they have less far to propagate in order to reach the hard spectral region,
thus giving shorter lags (\cite{Lyubarskii1997, Kotov2001, Arevalo2006, Rapisarda2016}).
The high/soft state in Cyg~X-1 has frequency dependent time lags
in its fast variability (\cite{Pottschmidt2000}), 
so must have radially stratified Compton continuum spectra 
rather than a single non-thermal Compton component. 

Thus the fast time lags strongly support radial stratification of the Comptonisation spectrum 
from Cyg X-1 rather than a single Comptonisation region. 
Here we critically reassess the spin of the high/soft state in Cyg X-1
by combining the complex Comptonisation continuum models with the best 
current models of the disk spectrum. 
Additionally, we apply these to the new data from Suzaku 
where the high energy Compton tail is the weakest ever observed in Cyg~X-1, 
so should give the best view of the intrinsic disk spectrum. 
We find that a two zone Comptonisation model gives a better fit to the spectrum 
than a single zone model, and significantly reduces the derived black hole spin. 

\section{Observations and Data Reduction}

\begin{figure}[htbp]
\centering
\includegraphics[width=14cm]{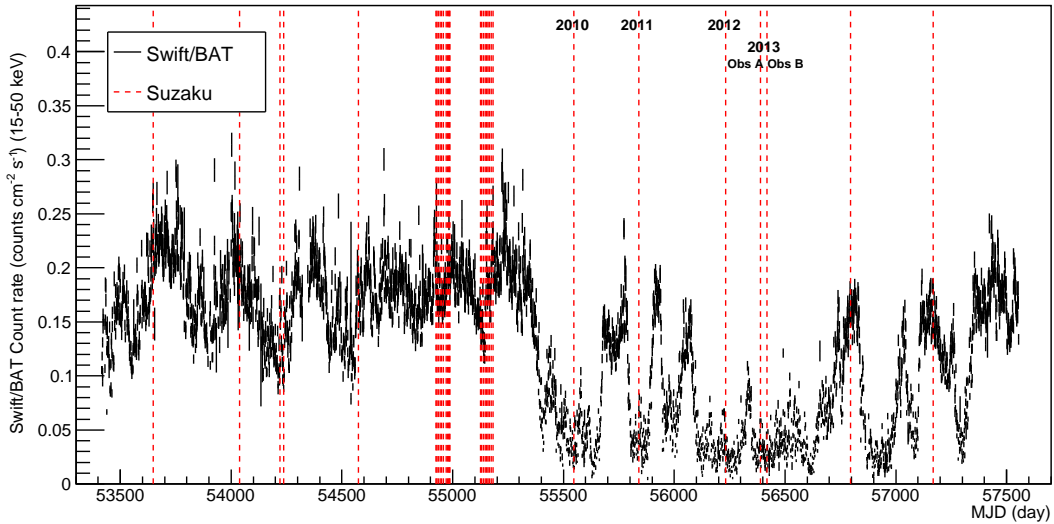} 
\caption{
Light curve of Cyg~X-1 observed by Swift/BAT.
Red dashed lines show the times of Suzaku observations.
There are five Suzaku observations in the high/soft state (defined as BAT count rate $<$ 0.05 counts cm$^{-2}$ s$^{-1}$).
}
\label{fig:lc_bat}
\end{figure}

Figure~\ref{fig:lc_bat} shows the Swift/BAT light curve of Cyg X-1\footnote{http://swift.gsfc.nasa.gov/results/transients/CygX-1}, 
with the times of all the Suzaku (\cite{Mitsuda2007}) observations superimposed.  
There are only five Suzaku observations in a high/soft state (BAT count rate of $< 0.05$), 
three of which have been already published 
(2010, 2011: \cite{Yamada2013}; 2012: \cite{Tomsick2014}).  
Here we focus on the remaining two pointed observations of the high/soft state of Cyg~X-1,
taken on 2013 April 8 (ObsID=407015010, hereafter Obs~A), 
and 2013 May 7 (ObsID=407015020, hereafter Obs~B).  

The XIS0 and XIS1 were operated with 1/4 window mode with 0.3-s burst option,
and XIS3 was operated with P-sum (fast timing: \cite{Koyama2007}).
We conducted the data reduction only for XIS0 as this is the best calibrated.
We excluded the core of the CCD image with a radius of 1.4~arcmin 
in order to reduce the pile-up fraction to below $\sim$3\% based on \citet{Yamada2012}. 
To estimate the background level, 
we analyzed the data during occultation with the GTI selection of $ELV<-5$.
The estimated XIS0 background is less than 0.01\% in the lower energy range ($\sim$2.0~keV) 
and less than $\sim$1\% in the higher energy range ($\sim$10~keV)
of the signal events. 
Hence we ignore this in our analysis. 

The HXD-PIN (\cite{Takahashi2007, Kokubun2007}) data reduction is 
conducted for both observations with the standard procedure, 
and the ftool of $hxdpinxbpi$ was used for tuned background files 
to create the PIN background spectrum, 
accounting for both the non-X-ray background (NXB) 
and the cosmic X-ray background (CXB) (\cite{Fukazawa2009}).

The exposure of PIN is 64.51~ks and 20.51~ks for Obs~A and Obs~B, respectively.
Because of the burst option, the corresponding effective exposure of XIS0 
is 6.61~ks and 1.39~ks for Obs~A and Obs~B, respectively. 
The combination of reduced effective exposure time from
excising the piled up core, and the 0.3-s burst option, 
means that we cannot constrain the fast time lags in the XIS0 data. 

\begin{figure}[htbp]
\begin{minipage}{0.5\hsize}
\centering
\includegraphics[width=8cm]{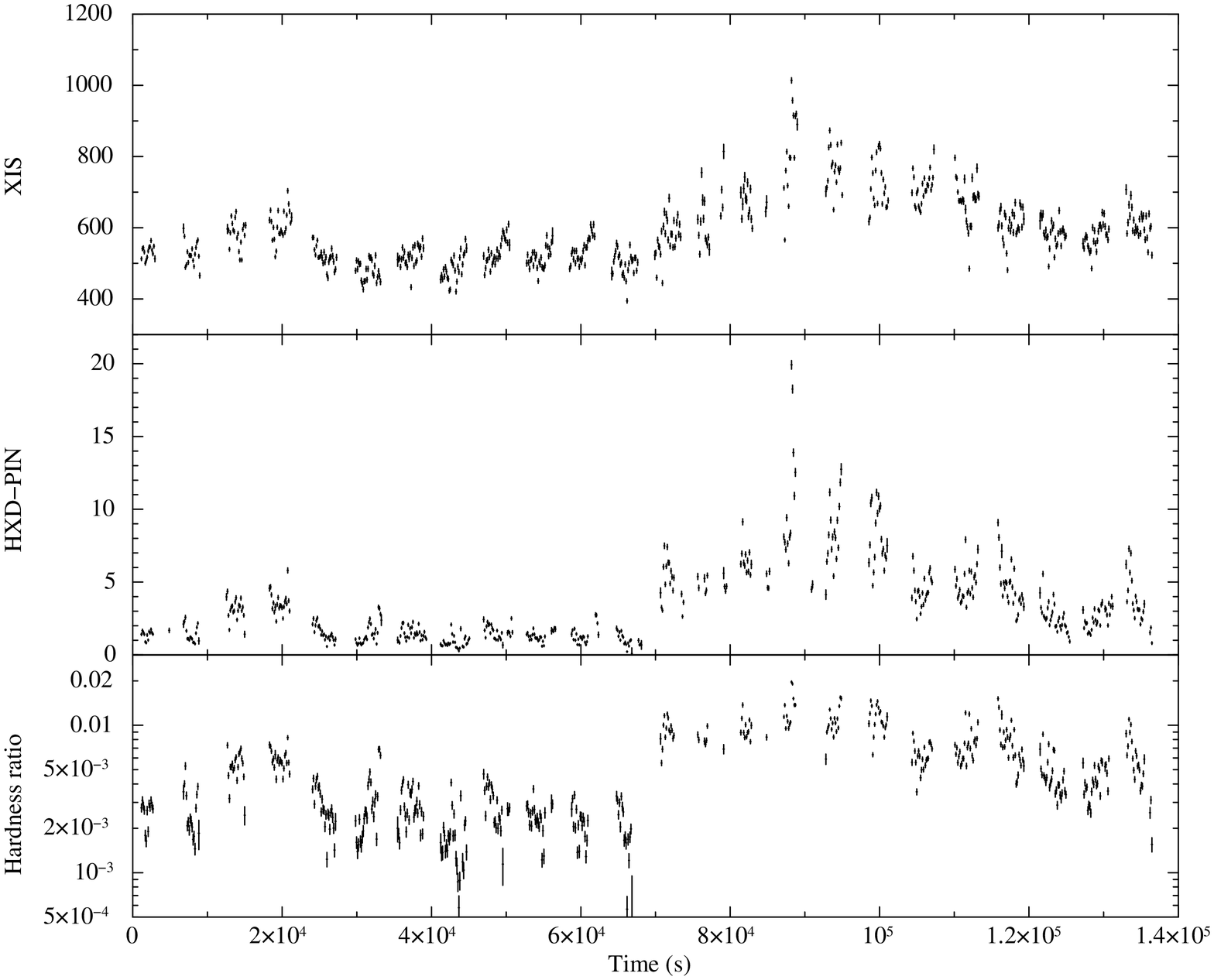} 
\end{minipage}
\begin{minipage}{0.5\hsize}
\centering
\includegraphics[width=8cm]{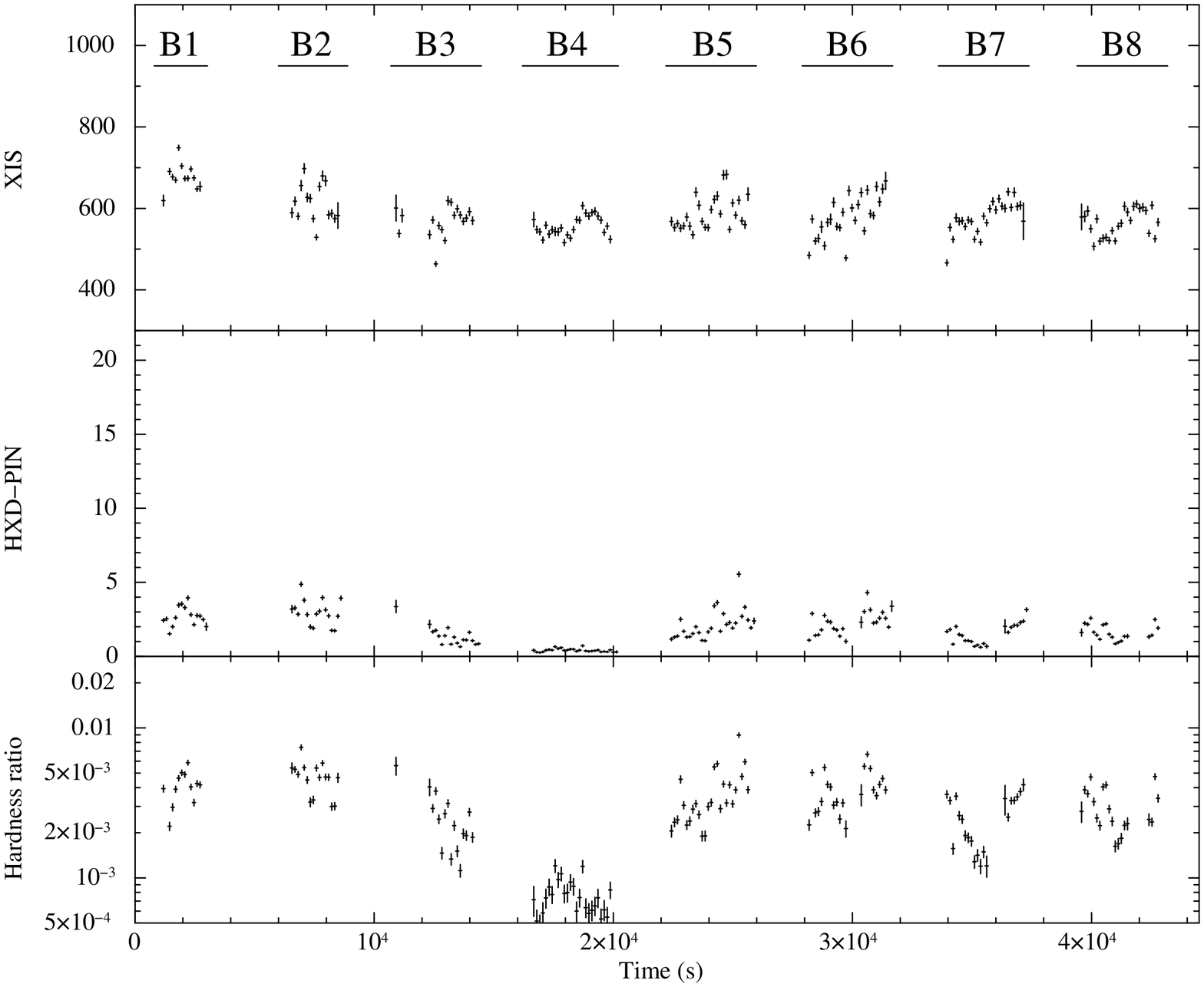}
\end{minipage}
\caption{
The light curves with bin of 128 sec observed by Suzaku during Obs~A (Left), and that of Obs~B (Right). 
Top: The light curve of the XIS in 0.5~--~10.0~keV.
Middle: The light curve of the PIN in 15~--30~keV. 
Bottom: The hardness ratio between PIN (15~--30~keV) and XIS (0.5~--~10.0~keV).
Obs~B4 has the lowest 15-30~keV flux and smallest hardness ratio of any orbital segment in Obs~A or B.
}
\label{fig:lc}
\end{figure}

Figure~\ref{fig:lc} shows the 0.5-10~keV XIS0 and 15-30~keV PIN light
curves and their hardness ratios for Obs~A and Obs~B. Plainly there
are spectral changes within these two observations.  In Obs~A there
are a few short excursions to very low hardness ratios, but in Obs~B
this low hardness is sustained over an entire orbit (the fourth
segment of Obs~B, hereafter Obs~B4).  Due to the pile up losses, we need
at least an orbit of data to accumulate enough signal-to-noise for a
good spectrum. Obs~B4 has the lowest hardness ratio of any of the orbit 
segments in Obs~A or Obs~B. 

\begin{figure}[htbp]
\begin{minipage}{0.5\hsize}
\centering
\includegraphics[width=8cm]{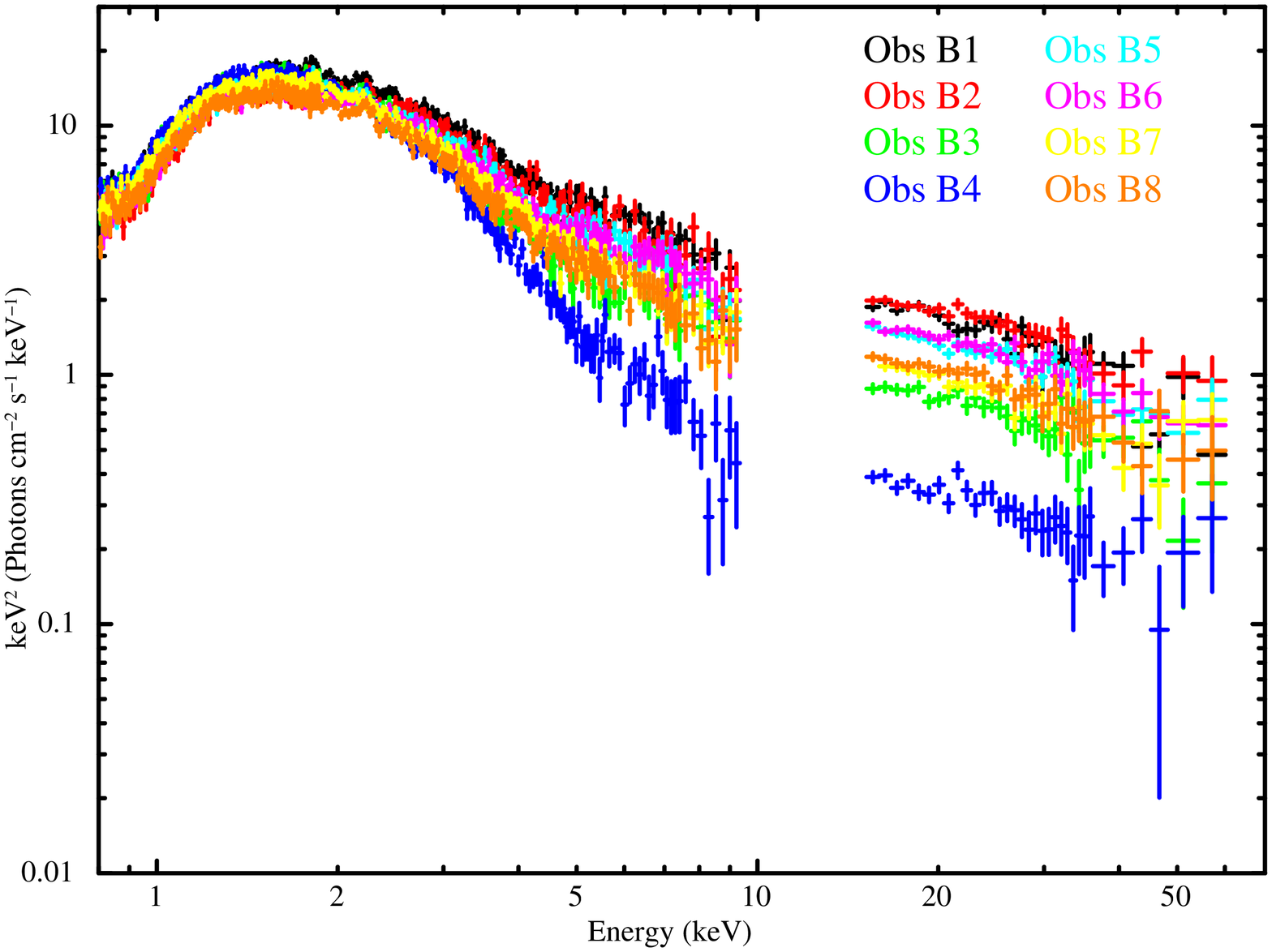} 
\end{minipage}
\begin{minipage}{0.5\hsize}
\centering
\includegraphics[width=8cm]{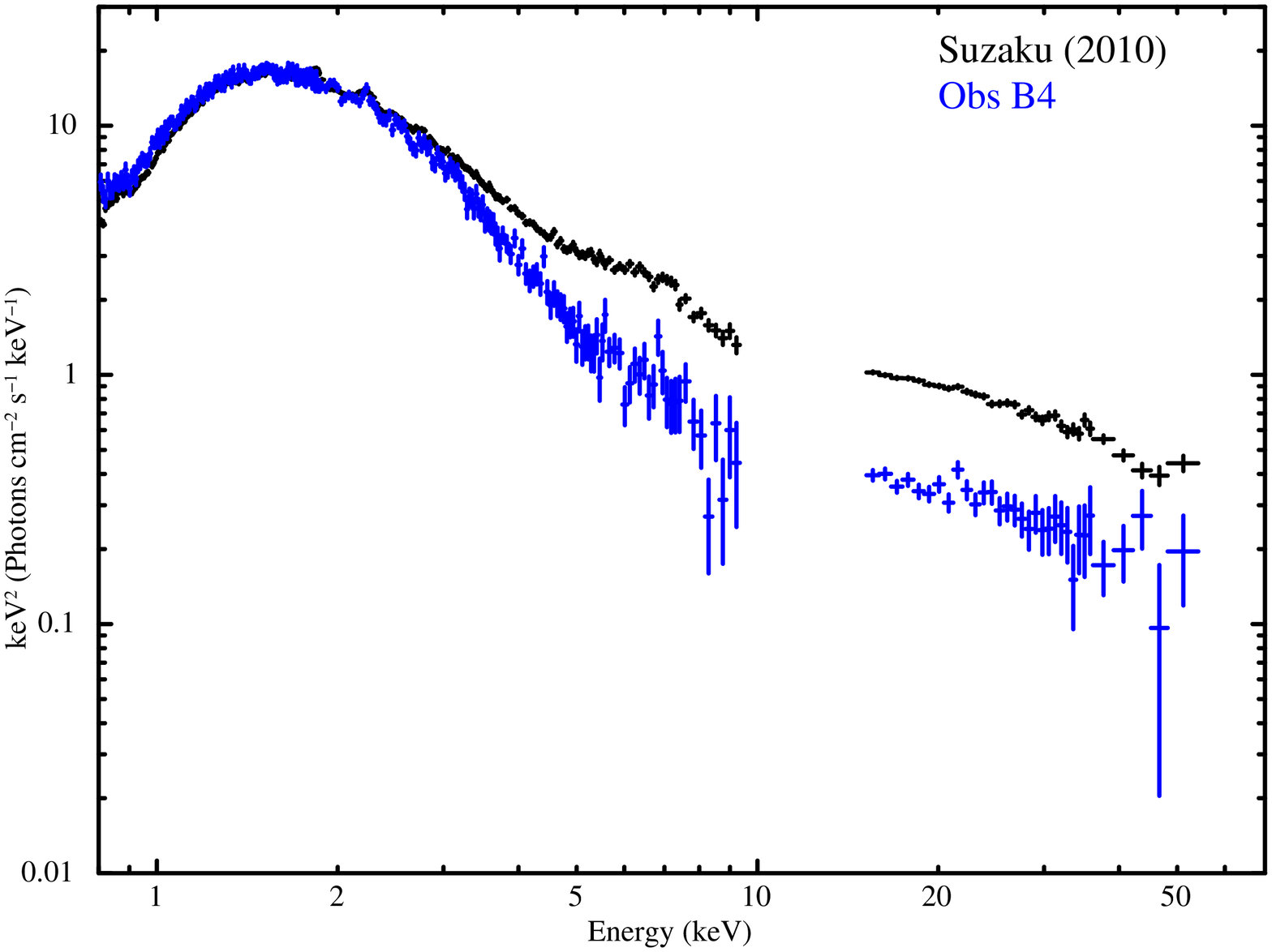}
\end{minipage}
\caption{
Left: The XIS, PIN spectrum during Obs~B1 (black), Obs~B2 (red), Obs~B3 (green), and Obs~B4 (blue), Obs~B5 (light blue), Obs~B6 (magenta), Obs~B7 (yellow), and Obs~B8 (orange), removing the instrumental responses. 
Right: The comparison between 2010 data (black) of Suzaku (used in \cite{Gou2014}) and Obs~B4 (blue).
}
\label{fig:spec_ab}
\end{figure}

To investigate the short-term ($\sim$3~ks) variation during Obs~B, 
we extract spectra from each orbit segment.
We give details of the time selection in Table~\ref{tab:observe}. 
Figure~\ref{fig:spec_ab} left panel shows the time averaged spectra of each of
the segments from Obs~B, 
unfolded with a single $\Gamma=2$ power-law model. 
It is clear that the 5 - 60 keV flux during Obs~B4 is the smallest, as
seen in the hardness ratio in Figure~\ref{fig:lc}. 
Figure~\ref{fig:spec_ab}  right panel shows Obs~B4 compared to a compilation of 
the previous softest spectra seen from Cyg~X-1, 
again unfolded assuming a single $\Gamma=2$ power-law model. 
The previous softest spectra seen from Cyg~X-1 by Suzaku was from 2010
(\cite{Yamada2013, Gou2014, Tomsick2014}), 
which is observed by the XIS0 operated with 1/4 window mode with 0.3-s burst option, 
where we reduced the data in same way as Obs~A and Obs~B.
\citet{Gou2014} compiled the softest Cyg~X-1 spectra 
where there were both CCD and higher energy data 
and showed that the Suzaku 2010 data were the softest ever seen at that point.
Since Obs~B4 is significantly softer, 
this means it is the softest spectrum yet observed 
for which there are both CCD data to constrain the disk 
and higher energy data to constrain the tail. 

There are more pointed RXTE observations than CCD datasets. 
\citet{Grinberg2014} uniformly analyzed all the archival RXTE data, 
characterizing their spectra 
with a phenomenological model of a disk plus broken power law with exponential cutoff.
We use their model and fit this to Obs~B4 over their 3-100~keV energy range. 
This gives a low energy spectral index of $\Gamma_1\sim 4.0$, 
steeper than their steepest index of $\Gamma_1\sim 3.63$ (\cite{Grinberg2014}).
Thus Obs~B4 is the softest spectrum yet detected from Cyg~X-1. 

\begin{table}[t]
\begin{center}
  \caption{The new observations of Cyg~X-1 in the high/soft state with Suzaku, together with details of the data extraction of the 2010 data used in Figure~\ref{fig:spec_ab} right.}\label{tab:observe}
\scalebox{0.8}[0.8]{ 
\begin{tabular}{rlllrr}
  \hline
  \hline
  Observation &Obs~ID &Obs~start &Obs~end  &\shortstack{Exposure [ks] \\ XIS } & \shortstack{Exposure [ks] \\ PIN}\\
  \hline
Obs~A  &407015010 &2013-04-08 02:17:18 &2013-04-09 23:00:19 &6.61 &64.51\\
Obs~B &407015020  &2013-05-07 02:13:17 &2013-05-07 14:34:19 &1.39 &20.51\\
B1&                            &2013-05-07 02:17:33 &2013-05-07 02:43:41 &0.19 &1.81\\
B2&                            &2013-05-07 03:47:49 &2013-05-07 05:04:13 &0.11 &2.10\\
B3&                            &2013-05-07 05:04:13 &2013-05-07 05:54:45 &0.13 &2.04\\
B4&                            &2013-05-07 06:37:01 &2013-05-07 07:30:23 &0.22 &3.45\\
B5&                            &2013-05-07 08:12:35 &2013-05-07 09:05:57 &0.19 &3.45\\
B6&                            &2013-05-07 09:48:11 &2013-05-07 11:27:33 &0.18 &2.97\\
B7&                            &2013-05-07 11:27:33 &2013-05-07 12:17:07 &0.19 &2.38\\
B8&                            &2013-05-07 12:59:23 &2013-05-07 13:52:39 &0.19 &2.31\\
2010      &905006010 &2010-12-17 00:10:26 &2010-12-17 18:41:36 &2.99 &35.50\\
\hline
\end{tabular}
}
\end{center}
\end{table}

\section{Data Analysis}

We use {\sc xspec} version 12.9.0n for all spectral fitting, 
and give 90\% confidence errors on all parameters ($\Delta\chi^2=2.706$). 
We fix the system parameters of distance, black hole mass and disk inclination 
at $D$ = 1.86~kpc, 
$M$ = 14.8~$M_{\odot}$,
and $i$ = 27.1~deg (\cite{Reid2011,Orosz2011}).
The solar abundance table is set to \citet{Wilms2000}, 
and iron abundance is set to 1.
We use data from 0.8-8~keV for XIS0, 
and 15-60~keV for the PIN, 
but we calculate all models over the range 0.1-1000 keV 
using the energies extend command
as they include convolution components. 
Additionally, in the following analysis, we fixed the relative
normalization {\sc const} of the XIS and PIN at
1.15\footnotemark\footnotetext{http://www.astro.isas.jaxa.jp/suzaku/doc/suzakumemo/suzakumemo-2008-06.pdf}.

\subsection{Disk with non-thermal Comptonisation}\label{sec:goumodel}

At first, we follow the model which is used in \citet{Gou2014}, 
where the spectrum is assumed to consist of a relativistic disk 
({\sc kerrbb}: \cite{Li2005}), 
with color temperature correction fixed to 1.7 (\cite{Shimura1995}), 
which acts as the source of seed photons for non-thermal Comptonization 
({\sc simpl}: \cite{Steiner2009}).  
The {\sc simpl} model in its original form only calculates the sum 
of the disk and Comptonised photons, 
whereas only the Comptonised photons are reflected.  
We thus use the extension of this model, 
{\sc csimpl} (\cite{Kolehmainen2014}) as it allows the Compton tail 
to be separated from the unscattered disk photons 
by setting the up scattering switch to 2.  
We denote this as {\sc csimpl(2)} (Compton scattering component), 
whereas {\sc csimpl(1)} is the sum of the unscattered disk
and Comptonised photons as in {\sc simpl}.  
\citet{Gou2014} used the {\sc xillver} tables of \citet{Garcia2014}
in order to describe the reflection from the disc.
These have the most up-to-date treatment of atomic physics, but 
assume that the illuminating spectrum is a power law,
i.e. that source of seed photons for Compton scattering is outside
of the bandpass. 
This is not true in our data.  
Instead we use the {\sc xilconv} model 
which converts the {\sc xillver} reflection table models into a convolution 
for use with any continuum (\cite{Kolehmainen2011}), parameterized by 
the ionization parameter $\xi=L/(nR^2)$ of the illuminated disc.
We then convolved this with the
relativistic smearing model {\sc relconv} (\cite{Dauser2014}) in order
to model reflection from the disc.
\citet{Gou2014} used {\sc tbabs} (\cite{Wilms2000}) 
for the absorption model, 
however we use {\sc tbnew\_gas}, which is an updated
version\footnotemark\footnotetext{http://pulsar.sternwarte.uni-erlangen.de/wilms/research/tbabs/}.
The full model is then given by:

\begin{description}
 \item[disk and Comptonization of disk photons (Standard model)]\mbox{}\\ 
CONST $\ast$ TBNEW\_GAS(CSIMPL(1) $\otimes$ KERRBB + RELCONV $\otimes$ XILCONV(CSIMPL(2) $\otimes$ KERRBB))
\end{description}

We initially tried to determine the best fit parameters of the relativistic reflection from the data, 
but there are too many free parameters 
(potentially three for a broken power law emissivity in {\sc relconv}, 
then another three which are the relative reflection normalization, 
iron abundance and ionization parameter in {\sc xilconv}) 
for the data quality. 
We follow \citet{Gou2014} and fix the radial illumination emissivity 
to a single power law of index of $2.5$, 
fix the abundance to solar, 
set the reflector solid angle to unity, 
the inner radius of {\sc relconv} is set to unity, i.e. tied to the spin dependent innermost stable circular orbit,
and leaving only $log(\xi)$ as a single free parameter. 
We note that if we leave additional parameters free,
their values cannot be constrained
and the fit is not improved.

\begin{figure}[htbp]
\centering
\includegraphics[width=8cm]{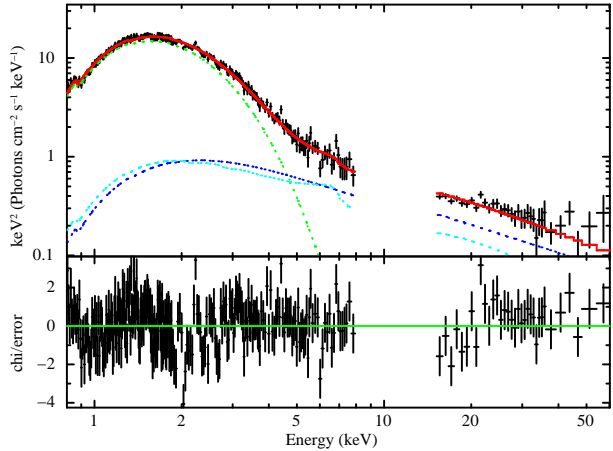} 
\caption{
The fitting results of Obs~B4. 
Assuming Standard disk model (disk and Comptonization of disk photons, corresponding to \cite{Gou2014}), 
red: total, green: {\sc kerrbb} , blue: Comptonization of {\sc kerrbb} , cyan: reflection of Comptonization.
}
\label{fig:fit_gou}
\end{figure}

The fitting results are shown in Figure~\ref{fig:fit_gou}, 
with parameters given in Table~\ref{tab:fit}.
This model gives a high spin, $a_*\sim 0.95$, 
similar to the results of \citet{Gou2014} even though 
the scattered fraction is significantly lower than in their data. 
While the fit is adequate, the PIN residuals in Fig.\ref{fig:fit_gou} have some discrepancies
below 20~keV, 
showing that the data are not well described by the high energy tail used in this model. 
A flatter spectral index would give a better fit at these energies, 
but the steeper index is required in order to fit the lower energy emission. Allowing 
more parameters to be free does not improve the fit, and as described above the values
for the extra parameters cannot be constrained. Instead, the discrepancies likely indicate
that the data would be better fit by two different Comptonisation components 
 (\cite{Gierlinski1999, Zdziarski2002}). Such a scenario is also more compatible
with the observed time lags, and we therefore explore more complex Comptonisation 
models, where there are two components.

\subsection{Complex Comptonisation}\label{sec:relativ}

We add in an additional thermal Comptonisation component
(hereafter called the +thermal model). 
This is a fairly good approximation 
(\cite{Gierlinski2003})
to the combined thermal and non-thermal
Comptonisation model used by 
\citet{Gierlinski1999} to fit the ASCA and RXTE soft state data from
Cyg~X-1 (which is only slightly harder than the Suzaku 2010 data).
However, their model was a single zone Comptonisation model,
where both thermal and non-thermal electrons co-exist in the same
region. Instead, our +thermal model explicitly has two separate Comptonisation 
components, which can be radially separated as required to explain the 
time lags.

We assume that the low temperature 
Comptonisation region is itself the seed photons for the non thermal
corona (see below), and use {\sc comptt} to describe the continuum shape
from Comptonization of soft photons \citep{Titarchuk1994}.  
We reflect both Comptonisation components, 
as we are assuming that the thermal electrons can also illuminate the disc. 
We approximate the disc itself by {\sc kerrbb} as before. 

We note that {\sc kerrbb} does not fully capture all the physics of
our physical picture, where there is a disk which extends down close
to (but not necessarily reaches) the last stable circular orbit of the
black hole, and where the disk has a ‘standard’ color temperature
correction only down to some radius, below which the color temperature
correction is significantly higher, effectively giving rise to the low
temperature, additional Comptonised emission. {\sc kerrbb} instead
assumes that the disc extends down to the last stable circular orbit,
with fixed color temperature correction. The {\sc kerrbb} models will then 
contain more high energy emission than a standard disc which 
truncates above the last stable circular orbit (leading to an underestimate of black hole spin).
However, the gravitational energy released below this truncation should power the
low temperature Comptonisation. Including this power in the {\sc kerrbb} model would
increase the inferred disc luminosity, leading to an overestimate of black hole spin. 
Since these two effects have opposite effects, and since 
there are no models available which combine
inhomogeneous disc structure with fully general relativistic 
ray tracing (see \cite{Kubota2016} for an inhomogeneous disc structure with approximate 
relativistic effects), we simply note that deriving spin is complex and probably not 
very robust when the spectra are not dominated by a standard disc component. 

\begin{description}
 \item[disk and thermal/non-thermal Comptonization and reflection (+thermal)]\mbox{}\\ 
 CONST $\ast$ TBNEW\_GAS(KERRBB + CSIMPL(1) $\otimes$ COMPTT + RELCONV $\otimes$ XILCONV(CSIMPL(1) $\otimes$ COMPTT))
\end{description}

\begin{figure}[htbp]
\begin{minipage}{0.5\hsize}
\centering
\includegraphics[width=8cm]{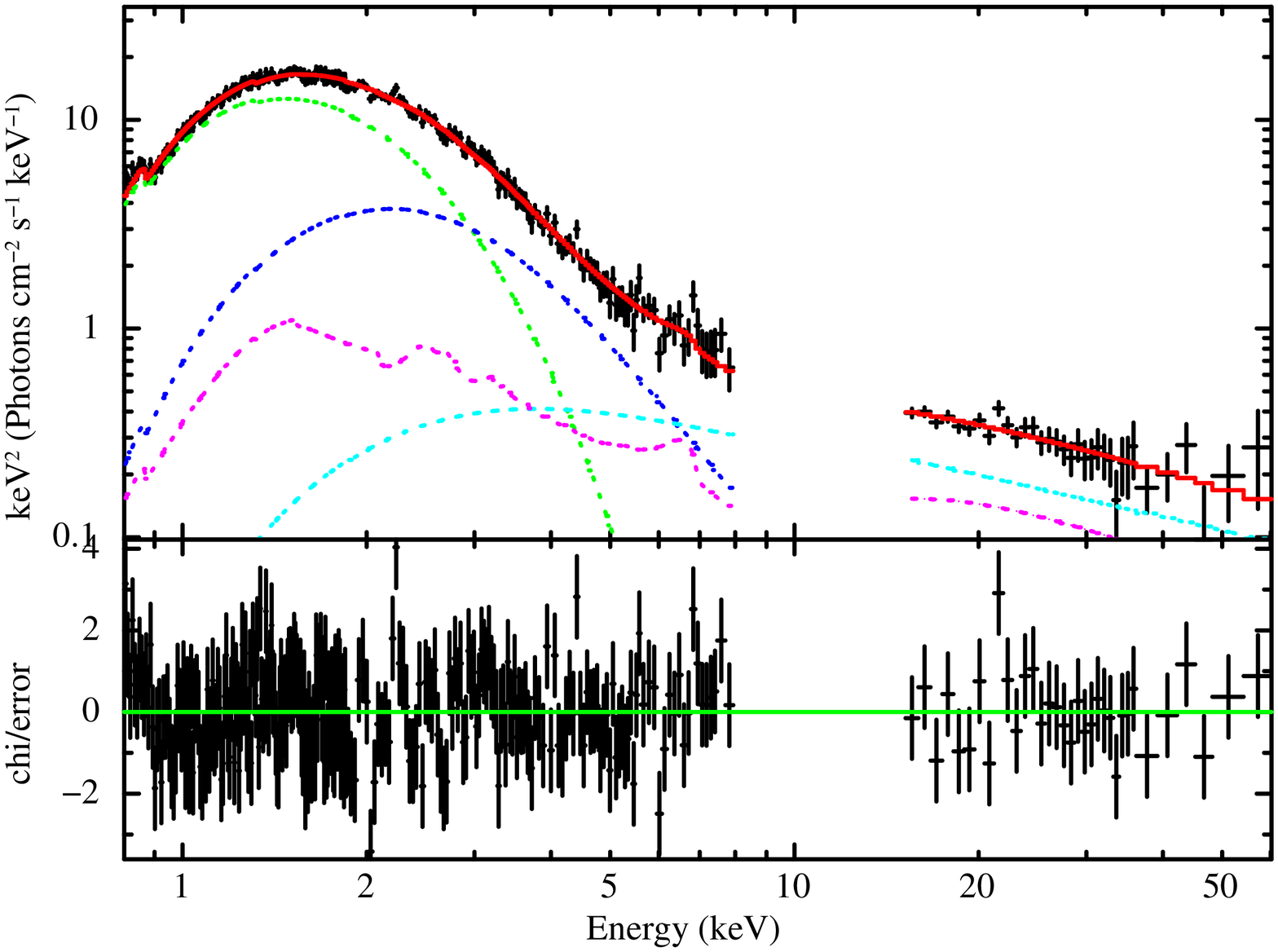} 
\end{minipage}
\begin{minipage}{0.5\hsize}
\centering
\includegraphics[width=8cm]{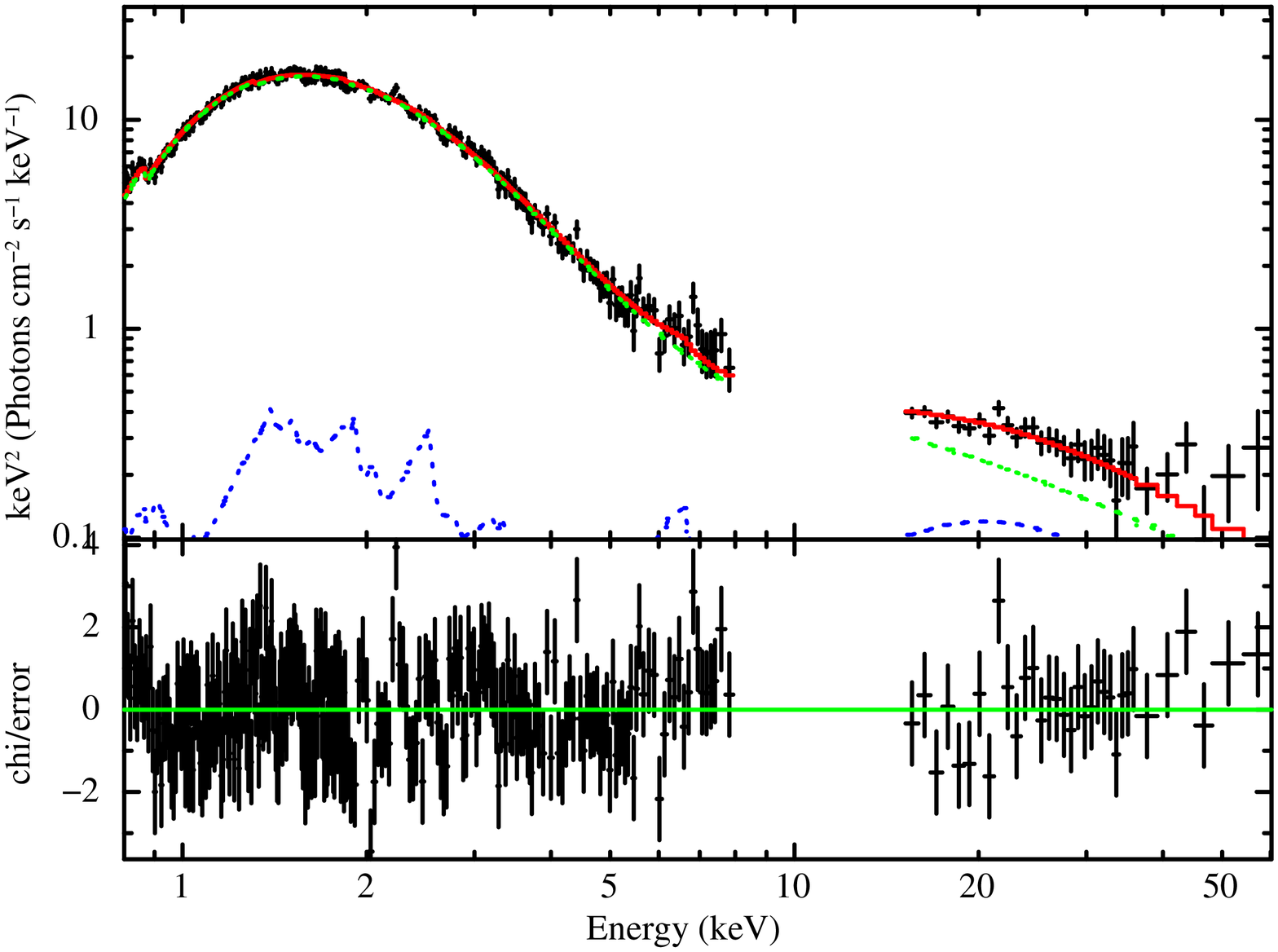}
\end{minipage}
\caption{
The fitting results of Obs~B4. 
Left: +thermal (disk and thermal/non-thermal Comptonization and reflection) model,
red: total, green: {\sc kerrbb} , blue: {\sc comptt} , cyan: Comptonization of {\sc comptt}, magenta: reflection of {\sc comptt} and Comptonization
Right: eqpair (non-relativistic disk with fully hybrid electrons and their reflection) model,
red: total, green: {\sc eqpair}, blue: reflection of {\sc eqpair}.
}
\label{fig:fit_comptt}
\end{figure}

This gives a significantly better fit to the data, 
at more than 5$\sigma$ confidence level by F-test (\cite{Protassov2002}),
and results in a significantly lower spin which is poorly constrained (See Table~\ref{tab:fit}).  
The best fit model is shown in Figure~\ref{fig:fit_comptt}.  
The emission which was previously fit 
as the highest temperature emission from the disk,
driving the high spin, 
is now fit instead as the Comptonization component (but note the caveats above on deriving spin). 

For completeness we also fit our data using the 
full model for thermal and non-thermal electrons used by 
\citet{Gierlinski1999}. This allows us to compare to previous 
results where the complex continuum was fit by a single-zone 
Comptonization model. 
We use the {\sc eqpair}\footnotemark\footnotetext{http://www.astro.yale.edu/coppi/eqpair/eqpap4.ps}, 
model, which has been used for both black hole binaries and AGN 
(e.g. \cite{Zdziarski1996, Zdziarski1998}).  
It is a sophisticated physical model which takes into account multiple 
heating and cooling processes in a single zone region.  
This derives the self consistent electron distribution from 
acceleration/heating balanced with cooling from both Coulomb collisions and Comptonisation. 
Coulomb collisions act to thermalize the low energy electrons, 
whereas Comptonisation cooling preserves a power law electron distribution. 
Thus the steady state electron distribution is hybrid, 
being thermal at low energies but with a power law tail 
due to Compton cooling dominating at high energies
(\cite{Poutanen1998, Coppi1999, Zdziarski2001, Hjalmarsdotter2016}).

We assume completely non-thermal acceleration, 
described by a single power law injected electron distribution 
with index $\Gamma_{\rm inj}$.  
We allow this index to be free, 
and fix the lower and upper limits of the electron Lorentz factors 
to the default values of $1.3$ and $1000$, respectively. 
Coulomb cooling depends on the density of electrons in the region
which is proportional to the optical depth, 
$\tau_{\rm p}$, divided by the size scale. 
We fix the size scale to the default of
$R=10^7$~cm (around $10R_g$, see \cite{Gierlinski1999}), 
but have $\tau_{\rm p}$ be a free parameter.  
Compton cooling depends on the density of seed photons, 
so is $\propto L_{bb}$ divided by the size scale, 
which is parameterized via the soft seed photon compactness,
$\ell_{bb}\propto L_{bb}/R =10$ (fixed, see \cite{Gierlinski1999}).
The seed photon distribution in {\sc eqpair} can be either a blackbody, 
or a simple Schwarzschild black hole model ({\tt diskpn}: \cite{Gierlinski1999}). 
We use the {\tt diskpn} option, which has two free parameters, 
namely the peak disk temperature (kT$_{\rm bb}$) and intrinsic normalization (norm), 
which is used as the overall normalization of {\sc eqpair}.

The full model is:
\begin{description}
\item[non-relativistic disk with fully hybrid electrons and their reflection (eqpair)]\mbox{}\\                                                                                           
 CONST $\ast$ TBNEW\_GAS(EQPAIR + RELCONV $\otimes$ XILCONV $\otimes$ EQPAIR)
\end{description}

The {\sc eqpair} model returns both the Comptonisation resulting from the hybrid electron
distribution and the unComptonised seed photons.  
However, our fits have a fairly high optical depth, 
so only a fraction $\exp(-\tau_{\rm p})$ of these unComptonised seed photons escape. 
Thus the spectrum is mainly the Comptonisation alone, so 
we reflect the entire {\sc eqpair} continuum 
to get a good approximation to the reflection 
from the Comptonisation components.  
This gives a slightly worse $\chi^2$ value than the +thermal model above, 
but is more constrained (has three fewer free parameters) so this is not significant. 
The inferred intrinsic {\tt diskpn} component has a very similar spectral shape 
to the {\sc kerrbb} component in the +thermal model, 
and the complex {\sc eqpair} Comptonisation is very similar 
in shape to the sum of the separate thermal and non-thermal Comptonisation components 
in the +thermal model. 
However, while this shows that a single zone model can reproduce most of the complex curvature, 
we again stress that a single-zone model can not explain the time lag behavior 
(\cite{Pottschmidt2000, Grinberg2014}).

\begin{table}[t]
\begin{center}
  \caption{Results of fitting the spectrum of B4}\label{tab:fit}
\scalebox{0.8}[0.8]{ 
\begin{tabular}{lcccc}
  \hline              
  \hline              
Component        & Parameter                                  &Standard                                  & +thermal                         & eqpair \\
  \hline
TBNEW\_GAS  &N$_{H}$ [10$^{22}$ cm$^{-2}$] &0.65$_{-0.02}^{+0.03}$          &0.68$_{-0.03}^{+0.01}$       &0.69$_{-0.02}^{+0.02}$\\
CSIMPL             &$\Gamma$                                &2.93$_{-0.05}^{+0.11}$            &2.67$_{-0.16}^{+0.18}$      &---\\
                          &FracSctr                                     &0.028$_{-0.003}^{+0.009}$     &0.07$_{-0.02}^{+0.06}$      &---\\
KERRBB\footnotemark[1] &eta                              &0 (fixed)                                   &0 (fixed)                              &---\\
                          &a$_*$                                         &0.95$_{-0.01}^{+0.01}$           &0.80$_{-0.30}^{+0.08}$      &---\\
                          &Mdd [g s$^{-1}$]                        &0.18$_{-0.01}^{+0.01}$            &0.24$_{-0.03}^{+0.06}$     &---\\
                          &rflag                                           &1 (fixed)                                    &1 (fixed)                             &---\\
                          &lflag                                           &0 (fixed)                                    &0 (fixed)                             &---\\
COMPTT\footnotemark[2] &T$_0$ [keV]              & ---                                            &0.44$_{-0.04}^{+0.16}$     &---\\
                          &kT [keV]                                      & ---                                            &3.52$_{-1.65}^{+28.73}$   &---\\
                          &taup                                            & ---                                             &1.66$_{-0.62}^{+10.99}$  &---\\
                          &norm                                           & ---                                             &1.37$_{-0.87}^{+1.63}$    &---\\
XILCONV          &log($\xi$)                                    &$>$3.79                                    &4.00$_{-0.63}^{+0.27}$     &4.18$_{-0.47}^{+0.28}$\\
EQPAIR\footnotemark[3] &l$_{\rm h}$/l$_{\rm s}$ & ---                                             & ---                                     &0.11$_{-0.01}^{+0.02}$\\
                         &l$_{\rm bb}$                                & ---                                             & ---                                      &10 (fixed)\\
                         &kT$_{\rm bb}$ [keV]                    & ---                                             & ---                                      &0.39$_{-0.02}^{+0.01}$\\
                         &l$_{\rm nt}$/l$_{\rm h}$               & ---                                             & ---                                      &1 (fixed)\\
                         &$\gamma_{\rm min}$                  & ---                                             & ---                                      &1.3 (fixed)\\
                         &$\gamma_{\rm max}$                 & ---                                             & ---                                      &1000 (fixed)\\
                         &$\Gamma_{\rm inj}$                    & ---                                             & ---                                      &5.0$_{-0.3}^{+0.0}$$^{\dagger}$\\
                         &radius [cm]                                  & ---                                             & ---                                      &10$^7$ (fixed)\\
                         &$\tau_{\rm p}$                             & ---                                             & ---                                      &1.51$_{-0.50}^{+1.48}$\\
                         &norm                                           & ---                                             & ---                                      &2.76$_{-0.45}^{+0.51}$\\
                         &$\chi^{2}$/d.o.f.                          &364.57/288                                &320.63/284                         &328.88/287\\
\hline
\multicolumn{5}{l}{Standard disk model: disk and Comptonization of disk photons}\\
\multicolumn{5}{l}{+thermal model: disk and thermal/non-thermal Comptonization and reflection}\\
\multicolumn{5}{l}{eqpair model: non-relativistic disk with fully hybrid electrons and their reflection}\\
%\footnote[1]{5}{rflag and lflag are fixed to the default values.}\\
%\footnote[2]{5}{assuming the geometry of the disk.}\\
%\footnote[3]{5}{l$_{\rm bb}$, l$_{\rm nt}$/l$_{\rm h}$, $\gamma_{\rm min}$, $\gamma_{\rm max}$, and the radius of the scattering region are fixed to the default values.}\\
\multicolumn{5}{l}{$^{1}$eta, rflag (a self-irradiation switch flag) and lflag (limb-darkening switch flag) }\\
\multicolumn{5}{l}{are fixed to the default values.}\\
\multicolumn{5}{l}{$^{2}$we fix the geometry switch at the default of 1 (disk).}\\
\multicolumn{5}{l}{$^{3}$the internal eqpair reflection is fixed at zero, and all parameters of eqpair associated }\\
\multicolumn{5}{l}{with this internally generated reflection model are frozen to their default values }\\
\multicolumn{5}{l}{so that we can use the newer reflection models of {\sc xilconv}. }\\
\multicolumn{5}{l}{See the text for parameter definitions.}\\
\multicolumn{5}{l}{$^{\dagger}$the upper limit of $\Gamma_{\rm inj}$ is pegged to 5.0.}\\
\hline
\end{tabular}
}
\end{center}
\end{table}

\section{Discussion and Conclusions}

The 2013 Suzaku data include the softest ever spectrum seen from Cyg~X-1.  
This means that these data have the smallest contamination 
from the high energy Compton tail, 
so should give the best determination of black hole spin from disk continuum fitting. 
However, we see clear evidence (more than 5$\sigma$ of F-test)
that there is additional low energy Comptonisation. 
Including this component significantly lowers the derived black hole spin compared 
to standard models where there is only a disk and high energy Compton tail.

Previous modeling of this low temperature thermal Comptonisation component was done using 
a single zone hybrid (thermal plus non-thermal) Comptonisation, 
which naturally produces low temperature thermal Comptonised emission 
even if the electrons are accelerated by purely non-thermal processes. 
These self consistent models derive the steady state electron distribution 
by balancing heating/acceleration with Coulomb and Compton cooling 
(\cite{Coppi1999}, \cite{Gierlinski1999}). 
However, the fast time lags in these data (\cite{Pottschmidt2000, Grinberg2014}) 
require that the Comptonisation is radially stratified, rather than arising from 
a single zone model. 
Then it seems more plausible that there are really two spatially separated electron distributions, 
with the low temperature thermal Comptonisation produced at larger radii, 
and non-thermal Comptonisation 
(or more likely, a true hybrid Comptonisation component) 
produced closer to the black hole
(see also \cite{Gierlinski2003, Hjalmarsdotter2016, Kubota2016}).

One possibility for the origin of the low temperature Comptonisation region 
is that it could be from the disk itself. 
Cyg~X-1 has a long term mass accretion rate 
which is close to the major hard-to-soft spectral transition, 
and it is not at all clear 
whether the source ever fully reaches the classic high/soft state seen in other, 
brighter, black hole binaries.
Other low inclination systems such as GX~339$-$4 and XTE~J1817$-$330 show 
ultra soft disk dominated spectra in their lowest luminosity soft states, 
where the disk dominated spectra have the lowest temperature. 
In Cyg~X-1 the ratio of unabsorbed 6-10~keV/2-6~keV flux in our softest spectrum 
is 0.05 not $<0.03$ as observed in GX~339$-$4 (\cite{MunozDarias2013}). 
Either Cyg~X-1 does not quite reach the soft state proper, 
or the spin in Cyg~X-1 is much higher than that of GX~339$-$4. 
We note that reflection fits generally give an extremely high spin for GX~339$-$4 
(e.g. \cite{Ludlam2015} but see \cite{Kubota2016}), 
while continuum fitting allows low spin solutions (\cite{Kolehmainen2010}). 

In the context of the truncated disk model for the spectral
transitions (e.g. \cite{Esin1997}; the review by \cite{Done2007}), an
incomplete transition means that the truncated inner disk is still
reforming, condensing out of the hot inner flow (\cite{Meyer2007}).
Hence its vertical structure could be somewhat different than that of
a standard thin disk.  
In particular, any remaining large scale magnetic fields could give pressure support 
(see e.g. \cite{Sadowski2016} and references therein), 
perhaps producing a transition Comptonised region between a fully condensed, 
outer thin disk and the hot inner flow.

The spectrum we see could then be a consequence of the disk structure
in a failed hard-to-soft spectral transition.  Other failed
hard-to-soft (or very long intermediate state) transitions are
reviewed in e.g. \citet{Negoro2010, Negoro2014}.  Of these,
XTE~J1753$-$0127 also shows a similarly very soft Comptonised spectrum
appearing on a failed hard to soft transition at very low luminosity.
However, it showed more standard disk spectra during higher luminosity
failed hard to soft transitions (\cite{Shaw2016}), so it is not at all
clear what is the trigger for this different behavior. We also note that similarly
complex Comptonisation is also seen in the much higher luminosity system GRS~1915$+$105, 
where again its inclusion lowers the derived black hole spin
(compare \cite{McClintock2006} with \cite{Middleton2006}). 

While we cannot do a full spectral-timing analysis on our data due to
limitations of the instrument modes, we note that the previous
high/soft states with stronger tails used RXTE PCA data
(\cite{Gou2011, Gou2014, Tomsick2014, Walton2016}).  Thus the fast
time lags in these spectra can be used to determine the soft Compton
component contribution unambiguously. We strongly urge full
spectral-timing analyses wherever possible in order to get
independent constraints on continuum complexity.

\bigskip

%Acknowledgement should be placed at end of main text.
%(NOT after the Appendix.)
\chapter*{Acknowledgements}
\thispagestyle{empty}
We acknowledge earlier work on these Cyg~X-1 data by Akifumi Yoshikawa.
This work is supported by Japan Society for the Promotion of Science (JSPS).
CD acknowledges support from a JSPS Invitation Fellowship for research in Japan (long term) L16518, 
and STFC funding under grant ST/L00075X. 
S.Y is supported by Grant-in-Aid for JSPS Kakenhi 15H05438/15H00785.
MA is an International Research Fellow of the JSPS.
We thank the referee for their comments, which greatly improved the clarity of the paper.

\appendix
%\section{Method of .....}
%\section{Approximation of ...}
%\section*{Complete data}

%%%
% See the manual for the detail.
%%%


\begin{thebibliography}{}

\bibitem[Ar{\'e}valo \& Uttley(2006)]{Arevalo2006}
Ar{\'e}valo, P., \& Uttley, P.\ 2006, \mnras, 367, 801 

\bibitem[Churazov et al.(2001)]{Churazov2001} 
Churazov, E., Gilfanov, M., \& Revnivtsev, M.\ 2001, \mnras, 321, 759 

\bibitem[Coppi(1999)]{Coppi1999} 
Coppi, P.~S.\ 1999, High Energy Processes in Accreting Black Holes, 161, 375 

\bibitem[Coriat et al.(2012)]{Coriat2012}
Coriat, M., Fender, R.~P., \& Dubus, G.\ 2012, \mnras, 424, 1991 

\bibitem[Dauser et al.(2014)]{Dauser2014}
Dauser, T., Garc{\'{\i}}a, J., Parker, M.~L., Fabian, A.~C., \& Wilms, J.\ 2014, \mnras, 444, L100 

\bibitem[Done et al.(2007)]{Done2007} 
Done, C., Gierli{\'n}ski, M., \& Kubota, A.\ 2007, \aapr, 15, 1 

\bibitem[Done \& Gierli{\'n}ski(2005)]{Done2005} 
Done, C., \& Gierli{\'n}ski, M.\ 2005, \mnras, 364, 208 

\bibitem[Dunn et al.(2010)]{Dunn2010}
Dunn, R.~J.~H., Fender, R.~P., K{\"o}rding, E.~G., Belloni, T., \& Cabanac, C.\ 2010, \mnras, 403, 61 

\bibitem[Esin et al.(1997)]{Esin1997}
Esin, A.~A., McClintock, J.~E., \& Narayan, R.\ 1997, \apj, 489, 865 

\bibitem[Fukazawa et al.(2009)]{Fukazawa2009}
Fukazawa, Y., Mizuno, T., Watanabe, S., et al.\ 2009, \pasj, 61, S17 

\bibitem[Garc{\'{\i}}a et al.(2014)]{Garcia2014}
Garc{\'{\i}}a, J., Dauser, T., Lohfink, A., et al.\ 2014, \apj, 782, 76 

\bibitem[Gierli{\'n}ski et al.(1999)]{Gierlinski1999}
Gierli{\'n}ski, M., Zdziarski, A.~A., Poutanen, J., et al.\ 1999, \mnras, 309, 496 

\bibitem[Gierli{\'n}ski \& Done(2003)]{Gierlinski2003} 
Gierli{\'n}ski, M., \& Done, C.\ 2003, \mnras, 342, 1083 

\bibitem[Gierli{\'n}ski \& Newton(2006)]{Gierlinski2006} 
Gierli{\'n}ski, M., \& Newton, J.\ 2006, \mnras, 370, 837 

\bibitem[Gladstone et al.(2007)]{Gladstone2007}
Gladstone, J., Done, C., \& Gierli{\'n}ski, M.\ 2007, \mnras, 378, 13 

\bibitem[Gou et al.(2011)]{Gou2011}
Gou, L., McClintock, J.~E., Reid, M.~J., et al.\ 2011, \apj, 742, 85 

\bibitem[Gou et al.(2014)]{Gou2014} 
Gou, L., McClintock, J.~E., Remillard, R.~A., et al.\ 2014, \apj, 790, 29 

\bibitem[Grinberg et al.(2014)]{Grinberg2014} 
Grinberg, V., Pottschmidt, K., B{\"o}ck, M., et al.\ 2014, \aap, 565, A1 

\bibitem[Hjalmarsdotter et al.(2016)]{Hjalmarsdotter2016} 
Hjalmarsdotter, L., Axelsson, M., \& Done, C.\ 2016, \mnras, 456, 4354

\bibitem[Kokubun et al.(2007)]{Kokubun2007}
Kokubun, M., Makishima, K., Takahashi, T., et al.\ 2007, \pasj, 59, 53 

\bibitem[Kolehmainen \& Done(2010)]{Kolehmainen2010}
Kolehmainen, M., \& Done, C.\ 2010, \mnras, 406, 2206 

\bibitem[Kolehmainen et al.(2011)]{Kolehmainen2011} 
Kolehmainen, M., Done, C., \& D{\'{\i}}az Trigo, M.\ 2011, \mnras, 416, 311 

\bibitem[Kolehmainen et al.(2014)]{Kolehmainen2014}
Kolehmainen, M., Done, C., \& D{\'{\i}}az Trigo, M.\ 2014, \mnras, 437, 316 

\bibitem[Kotov et al.(2001)]{Kotov2001} 
Kotov, O., Churazov, E., \& Gilfanov, M.\ 2001, \mnras, 327, 799 

\bibitem[Koyama et al.(2007)]{Koyama2007}
Koyama, K., Tsunemi, H., Dotani, T., et al.\ 2007, \pasj, 59, 23 

\bibitem[Kubota \& Done(2016)]{Kubota2016} 
Kubota, A., \& Done, C.\ 2016, \mnras, 458, 4238

\bibitem[Li et al.(2005)]{Li2005} 
Li, L.-X., Zimmerman, E. R., Narayan, R., \& McClintock, J. E. 2005, \apjs, 157, 335

\bibitem[Ludlam et al.(2015)]{Ludlam2015}
Ludlam, R.~M., Miller, J.~M., \& Cackett, E.~M.\ 2015, \apj, 806, 262 

\bibitem[Lyubarskii(1997)]{Lyubarskii1997}
Lyubarskii, Y.~E.\ 1997, \mnras, 292, 679 

\bibitem[Maccarone \& Coppi(2003)]{Maccarone2003} 
Maccarone, T.~J., \& Coppi, P.~S.\ 2003, \mnras, 338, 189 

\bibitem[McClintock et al.(2006)]{McClintock2006} 
McClintock, J.~E., Shafee, R., Narayan, R., et al.\ 2006, \apj, 652, 518 

\bibitem[Meyer et al.(2007)]{Meyer2007} 
Meyer, F., Liu, B.~F., \& Meyer-Hofmeister, E.\ 2007, \aap, 463, 1 

\bibitem[Middleton et al.(2006)]{Middleton2006} 
Middleton, M., Done, C., Gierli{\'n}ski, M., \& Davis, S.~W.\ 2006, \mnras, 373, 1004 

\bibitem[Mitsuda et al.(2007)]{Mitsuda2007}
Mitsuda, K., Bautz, M., Inoue, H., et al.\ 2007, \pasj, 59, 1 

\bibitem[Miyamoto \& Kitamoto(1989)]{Miyamoto1989} 
Miyamoto, S., \& Kitamoto, S.\ 1989, \nat, 342, 773 

\bibitem[Mu{\~n}oz-Darias et al.(2013)]{MunozDarias2013} 
Mu{\~n}oz-Darias, T., Coriat, M., Plant, D.~S., et al.\ 2013, \mnras, 432, 1330 

\bibitem[Negoro et al. (2010)]{Negoro2010}
Negoro, H., \& Maxi Team 2010, The First Year of MAXI: Monitoring Variable X-ray Sources, 6 

\bibitem[Negoro et al. (2014)]{Negoro2014} 
Negoro, H., \& MAXI Team 2014, Suzaku-MAXI 2014: Expanding the Frontiers of the X-ray Universe, 128 

\bibitem[Nowak et al.(1999)]{Nowak1999}
Nowak, M.~A., Vaughan, B.~A., Wilms, J., Dove, J.~B., \& Begelman, M.~C.\ 1999, \apj, 510, 874 

\bibitem[Orosz et al.(2011)]{Orosz2011} 
Orosz, J.~A., McClintock, J.~E., Aufdenberg, J.~P., et al.\ 2011, \apj, 742, 84 

\bibitem[Pottschmidt et al.(2000)]{Pottschmidt2000} 
Pottschmidt, K., Wilms, J., Nowak, M.~A., et al.\ 2000, \aap, 357, L17 

\bibitem[Poutanen \& Coppi(1998)]{Poutanen1998}
Poutanen, J., \& Coppi, P.~S.\ 1998, Physica Scripta Volume T, 77, 57 

\bibitem[Protassov et al.(2002)]{Protassov2002}
Protassov, R., van Dyk, D.~A., Connors, A., Kashyap, V.~L., \& Siemiginowska, A.\ 2002, \apj, 571, 545 

\bibitem[Rapisarda et al.(2016)]{Rapisarda2016} 
Rapisarda, S., Ingram, A., Kalamkar, M., \& van der Klis, M.\ 2016, \mnras, 462, 4078 

\bibitem[Reid et al.(2011)]{Reid2011} 
Reid, M.~J., McClintock, J.~E., Narayan, R., et al.\ 2011, \apj, 742, 83 

\bibitem[Remillard \& McClintock(2006)]{Remillard2006}
Remillard, R.~A., \& McClintock, J.~E.\ 2006, \araa, 44, 49 

\bibitem[S{\c a}dowski(2016)]{Sadowski2016} 
S{\c a}dowski, A.\ 2016, \mnras, 462, 960 

\bibitem[Shaw et al.(2016)]{Shaw2016}
Shaw, A.~W., Gandhi, P., Altamirano, D., et al.\ 2016, \mnras, 458, 1636 

\bibitem[Shimura \& Takahara(1995)]{Shimura1995} 
Shimura, T., \& Takahara, F.\ 1995, \apj, 445, 780 

\bibitem[Smith et al.(2002)]{Smith2002}
Smith, D.~M., Heindl, W.~A., \& Swank, J.~H.\ 2002, \apj, 569, 362 

\bibitem[Steiner et al.(2009)]{Steiner2009}
Steiner, J.~F., Narayan, R., McClintock, J.~E., \& Ebisawa, K.\ 2009, \pasp, 121, 1279 

\bibitem[Sugimoto et al.(2016)]{Sugimoto2016}
Sugimoto, J., Mihara, T., Kitamoto, S., et al.\ 2016, \pasj, 68, S17 

\bibitem[Takahashi et al.(2007)]{Takahashi2007}
Takahashi, T., Abe, K., Endo, M., et al.\ 2007, \pasj, 59, 35 

\bibitem[Titarchuk(1994)]{Titarchuk1994}
Titarchuk, L.\ 1994, \apj, 434, 570 

\bibitem[Tomsick et al.(2014)]{Tomsick2014} 
Tomsick, J.~A., Nowak, M.~A., Parker, M., et al.\ 2014, \apj, 780, 78 

\bibitem[van Paradijs(1996)]{van1996} 
van Paradijs, J.\ 1996, \apjl, 464, L139 

\bibitem[Walton et al.(2016)]{Walton2016} 
Walton, D.~J., Tomsick, J.~A., Madsen, K.~K., et al.\ 2016, \apj, 826, 87 

\bibitem[Wilms et al.(2000)]{Wilms2000}
Wilms, J., Allen, A., \& McCray, R.\ 2000, \apj, 542, 914 

\bibitem[Yamada et al.(2012)]{Yamada2012} 
Yamada, S., Uchiyama, H., Dotani, T., et al.\ 2012, \pasj, 64, 53 

\bibitem[Yamada et al.(2013)]{Yamada2013} 
Yamada, S., Negoro, H., Torii, S., et al.\ 2013, \apjl, 767, L34 

\bibitem[Yu \& Yan(2009)]{Yu2009} 
Yu, W., \& Yan, Z.\ 2009, \apj, 701, 1940 

\bibitem[Zdziarski et al.(1996)]{Zdziarski1996}
Zdziarski, A.~A., Johnson, W.~N., \& Magdziarz, P.\ 1996, \mnras, 283, 193 

\bibitem[Zdziarski et al.(1998)]{Zdziarski1998}
Zdziarski, A.~A., Poutanen, J., Mikolajewska, J., et al.\ 1998, \mnras, 301, 435 

\bibitem[Zdziarski et al.(2001)]{Zdziarski2001}
Zdziarski, A.~A., Grove, J.~E., Poutanen, J., Rao, A.~R., \& Vadawale, S.~V.\ 2001, \apjl, 554, L45 

\bibitem[Zdziarski et al.(2002)]{Zdziarski2002}
Zdziarski, A.~A., Poutanen, J., Paciesas, W.~S., \& Wen, L.\ 2002, \apj, 578, 357 



\end{thebibliography}
\end{document}